    \documentclass[aps,prl,showpacs,amsmath,twocolumn]{revtex4}

\usepackage{graphicx}
\usepackage{dcolumn}
\usepackage{bm}
\usepackage{color}

\usepackage{amssymb}

\newcommand{\NS}{\text{NS}}

\newcommand{\thr}{\text{th}}

\begin{document}
\title{Non-Newtonian granular hydrodynamics. What do the inelastic simple shear flow and the elastic Fourier flow have in common?}
\author{Francisco Vega Reyes}
\author{Andr\'es Santos}
\author{Vicente Garz\'o}
\affiliation{Departamento de F\'{\i}sica, Universidad de Extremadura, E-06071
Badajoz, Spain}

\begin{abstract}
We describe a special class of steady Couette flows in dilute granular gases admitting a non-Newtonian hydrodynamic description for strong dissipation. The class occurs when viscous heating exactly balances inelastic cooling, resulting in a uniform heat flux. It includes the Fourier flow of ordinary gases and the simple or uniform shear flow (USF) of granular gases as special cases. The rheological functions have the same values as in the USF and generalized thermal conductivity coefficients  can be identified.  These points are confirmed by molecular dynamics simulations, Monte Carlo simulations of the Boltzmann equation, and analytical results from Grad's 13-moment method.
\end{abstract}


\date{\today}
\maketitle

The study of granular matter is interesting from a technological point view because {its understanding has many applications in technology and in sciences other than physics, such as biology \cite{AT06,UMS96}. Furthermore, it is also important from a more fundamental point of view \cite{Go03}. For example, the Boltzmann equation (BE) for low-density granular gases describes a generalization of the contraction from a microscopic to mesoscopic scale and thus a generalization of fundamental concepts in the fields of statistical and fluid mechanics. }The BE for granular gases (usually modeled as smooth inelastic hard spheres)
has been widely employed to analyze several granular flow problems and a large number of research works have been recently published in this field \cite{AT06,TG98}. A standard approach used for solving the BE for ordinary gases consists in obtaining a perturbative solution, which results in Navier--Stokes (NS) {or Burnett type} hydrodynamic equations \cite{Go03}.  However, the kinetic energy loss in the collisions renders the granular steady flows inherently non-Newtonian \cite{Go03,TTM01,SGD04}.

We report in this Letter on strong evidence of non-Newtonian hydrodynamic steady states  in the planar Couette flow geometry for a wide range of inelasticities. Furthermore, our description is inclusive in the sense that it comprises a class (manifold) of steady flows,  whose elements correspond both to granular and  ordinary gases.  This novel class  occurs when the heat flux $\mathbf{q}$ is  constant across the system, due to an exact \emph{local} balance of inelastic cooling and viscous heating, even though the temperature and the shear rate are in general not uniform. As a consequence, this class of nonlinear flows has several surprising hydrodynamic properties: (i) the nonlinear temperature profiles $T(y)$ are indistinguishable from those of the Fourier steady state of an ordinary gas with the same temperature difference (Fourier flows are the ``elastic-limit'' elements of the manifold); (ii) when the  spatial coordinate $y$ normal to the moving plates is eliminated between temperature and flow velocity the resulting profiles $T(u_x)$ are  \emph{linear};  (iii) the non-Newtonian rheological properties (shear stress and anisotropic temperatures) are uniform and have the same values as those obtained in the well known simple or uniform shear flow (USF) of granular gases \cite{Go03}, which is in fact the particular case $\mathbf{q}=\mathbf{0}$ of the manifold; and (iv) the heat flux vector  is proportional to the thermal gradient (generalized Fourier's law) with an effective thermal conductivity tensor. Because of property (ii), we will refer henceforth to these flows as ``linear $T(u_x)$ flows,'' or simply, ``LTu'' flows. Property (iv) can be interpreted as a method for measuring the intrinsic thermal conductivity coefficients of the USF state directly from LTu steady states.
All these results are supported by three independent and complementary routes: an approximate analytical solution from Grad's 13-moment {(G13)} method to the BE, direct simulation Monte Carlo (DSMC) numerical solutions of the BE, and molecular dynamics (MD) simulations.
{The existence of the special LTu class at NS order was theoretically proven  in a recent work \cite{VU09}, but the applicability of the NS description is restricted to the quasi-elastic limit  and so the general proof of LTu states requires a non-Newtonian description, as carried out in this Letter.}

Conservation of momentum in the steady state Couette flow implies $P_{iy}=\text{const}$, where $P_{ij}$ is the stress tensor. As for the energy balance equation, it   reads
\begin{equation}
{-{\partial_y q_y}=({3}/{2})nT\zeta+P_{xy}{\partial_y u_x},}
\label{Tbal}
\end{equation}
where $n$ is the number density and $\zeta$ is the inelastic cooling rate.
Equation \eqref{Tbal} is  valid for all steady states in the system, whether  hydrodynamics applies or not. As  can be seen, the signature of the heat flux gradient is determined by the balance between two terms: the first one on the right-hand side comes from inelastic cooling and is inherently positive; the second term is due to viscous heating  and is inherently negative. Thus, the condition for homogeneous heat flux is that these two terms  exactly balance in the whole bulk region. Moreover, the streamwise heat flux component $q_x$  (absent at NS order) turns out to be ``synchronized'' to the crosswise component $q_y$ in the sense that it becomes homogeneous when so does $q_y$.

\begin{figure}
 \includegraphics[width=0.75\columnwidth]{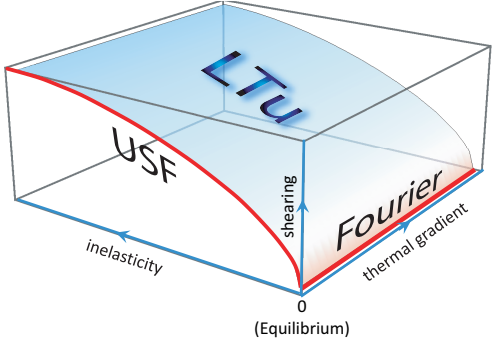}
\caption{(color online) Each point of this diagram represents a  Couette flow steady state. The  surface defines the LTu class, which contains the lines representing the Fourier flow for ordinary gases and the USF for granular gases.}\label{diagram}
\end{figure}

\begin{figure}
  \includegraphics[width=0.75\columnwidth]{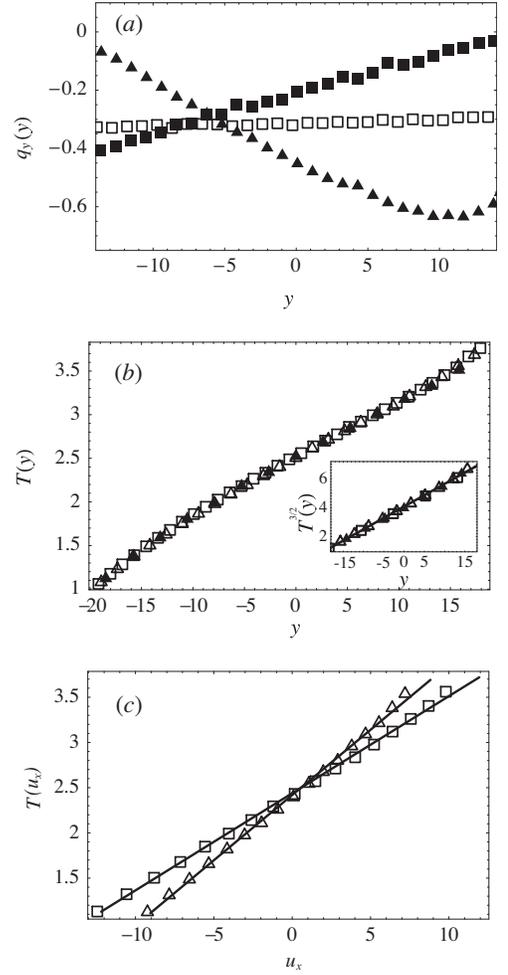}
\caption{(\textrm{a}) Profiles of the heat flux component $q_y$  for $\alpha=0.7$, $\Delta T=5$, and a shear rate 8\% smaller than the threshold value ($\blacktriangle$),  equal to the threshold value  ($\square$), and  4\% larger than the threshold value ($\blacksquare$). (\textrm{b}) Temperature profiles in the LTu flow for a common fluid temperature difference $T(h/2)-T(-h/2)=4$  and $\alpha=0.5$ ($\square$), $\alpha=0.7$ ($\triangle$), and $\alpha=1$ ($\blacktriangle$). In the latter case (ordinary gas), the simulated state is the conventional Fourier flow (without shearing). The inset shows  $T^{3/2}(y)$. (\textrm{c})  $T(u_x)$ profiles for the LTu class with $T(h/2)-T(-h/2)=4$  and $\alpha=0.5$ ($\square$) and $\alpha=0.7$ ($\triangle$). The shown data have been obtained by DSMC simulations.}\label{TigualT}
\end{figure}

Application of {the classical G13 method} \cite{G49} to the BE for inelastic collisions {yields
$\zeta=\frac{5}{12}\nu(1-\alpha^2)$ (where $\alpha$ is the coefficient of normal restitution and $\nu=\frac{16}{5}\sqrt{\pi}n\sigma^2\sqrt{T/m}$ is an effective collision frequency, $\sigma$ and $m$ being the diameter and mass of a sphere, respectively) and} a closed set of coupled equations for the hydrodynamic fields, the stress tensor, and the heat flux. This set  allows for an LTu solution characterized by $nT=\text{const}$,
\begin{equation}
\nu^{-1}\partial_y T=A=\text{const},\quad \nu^{-1}\partial_y u_x=a(\alpha)=\text{const},
\label{1}
\end{equation}
\begin{equation}
P_{xy}=-\eta(\alpha)\partial_y u_x=\text{const},\quad \frac{P_{ii}}{nT }=\theta_i(\alpha)=\text{const},
\label{2}
\end{equation}
\begin{equation}
q_y=-\lambda(\alpha)\partial_y T=\text{const},\quad q_x=\phi(\alpha)\partial_y T=\text{const}.
\label{3}
\end{equation}
{Since the trace of the stress tensor is $3nT$, one has $\theta_x+\theta_y+\theta_z=3$.}
Equation \eqref{1} implies that  $\partial T/\partial u_x=A/a(\alpha)$, so that $T$ is indeed a linear function of $u_x$.  Equations \eqref{1}--\eqref{3} define a class of solutions  because the constant $A$ is arbitrary. On the other hand, the dimensionless shear rate $a(\alpha)$ (which is the Knudsen number associated with the shearing), the  effective shear viscosity $\eta^*(\alpha)\equiv\eta(\alpha)/\eta_\NS(1)$,  the temperature ratios $\theta_i(\alpha)$, and the effective thermal conductivities $\lambda^*(\alpha)\equiv\lambda(\alpha)/\lambda_\NS(1)$ and $\phi^*(\alpha)\equiv\phi(\alpha)/\lambda_\NS(1)$  are independent of $A$, their values depending on inelasticity only. Here $\eta_\NS(1)=nT/\nu$ and $\lambda_\NS(1)=15\eta_\NS(1)/4m$ are the NS transport coefficients of the elastic gas \cite{G49}. The quantities $a(\alpha)$ and $\eta^*(\alpha)$ are related each other by the exact balance equation \eqref{Tbal} and the LTu condition $\partial_y q_y=0$,
\begin{equation}
  a^2(\alpha)=3\zeta^*(\alpha)/2\eta^*(\alpha),
  \label{4}
\end{equation}
where $\zeta^*(\alpha)\equiv \zeta/\nu$. Equation \eqref{4} shows that in the elastic limit $\alpha\to 1$ the LTu shear rate vanishes ($a\to 0$) and thus the conventional Fourier flow ($A\neq 0$) for an ordinary gas is included in the LTu class as a special case. Conversely,  the USF is recovered as another special case in the limit $A\to 0$ with $\alpha<1$. This is sketched in Fig.\ \ref{diagram}.
The explicit expressions predicted by {the G13} method for the transport coefficients are \cite{note}
\begin{equation*}
\eta^*=\frac{\beta_1}{(\beta_1+\zeta^*)^2}, \quad \theta_x=\frac{\beta_1+3\zeta^*}{\beta_1+\zeta^*},\quad \theta_y=\theta_z,
\end{equation*}
\begin{equation*}
\lambda^*=\beta_2(70\theta_y-20+63\zeta^*/\beta_2)/(50\beta_2^2-63a^2),
\end{equation*}
\begin{equation*}
\phi^*=7a(21\theta_y-6+10\beta_2\eta^*)/(50\beta_2^2-63a^2),
\end{equation*}
 where  $\beta_1(\alpha)=(1+\alpha)(2+\alpha)/6$ and $\beta_2(\alpha)=(1+\alpha)(49-33\alpha)/32$.
It must be noticed that $\eta^*(\alpha)\neq\eta^*_\NS(\alpha)=(\beta_1+\frac{1}{2}\zeta^*)^{-1}$ and $\lambda^*(\alpha)\neq\lambda^*_\NS(\alpha)=(\beta_2-3\zeta^*)^{-1}$.

To validate  {the G13} theoretical predictions, we have performed DSMC simulations of the BE and MD simulations ($\text{global solid fraction}=7\times 10^{-3}$) for a granular gas of hard spheres enclosed   between two plates located at $y=\pm h/2$ and moving with velocities $U_\pm$ (see Ref.\ \cite{VU09} for technical details). Diffuse boundary conditions characterized by wall temperatures $T_\pm$ ($T_-\leq T_+$) are applied. In what follows, quantities are nondimensionalized by the choice of units $m=1$, $T(-h/2)=1$, $n(-h/2)=1$, and  $\nu(-h/2)=1$. In these units, the quantity $A$ represents the \emph{maximum} value across the system of the Knudsen number associated with the thermal gradient \cite{VU09}. The separation between the plates has typically been set $h\approx 5$--$20$ and we have considered a wall temperature difference in the range $\Delta T\equiv T_+/T_--1=0$--$20$. We have looked for Couette flows belonging to the LTu class by fixing  $\Delta T$ and  varying the applied shear $\dot{\gamma}=(U_{+}-U_{-})/h$. Once the steady state is reached, we monitor the parametric plot of temperature versus flow field, $T(u_x)$. We have observed in all the cases  a definite sign of the curvature parameter $\partial^2 T/\partial u_x^2$ in the bulk, with no inflection point. Interestingly, as the shearing increases and  a certain threshold value $\dot{\gamma}=\dot{\gamma}_\thr$ is crossed, the sign of $\partial^2 T/\partial u_x^2$ undergoes a change from negative to positive. At the  threshold shear $\dot{\gamma}_\thr$, $\partial^2 T/\partial u_x^2=0$ and this signals the onset of the LTu flow, as explained above. This transition is accompanied by a change in the slope of $ q_y$, so one also has  $q_y=\text{const}$ at $\dot{\gamma}=\dot{\gamma}_\thr$. This is illustrated by DSMC data in Fig.\ \ref{TigualT}(a), while Figs.\ \ref{TigualT}(b) and \ref{TigualT}(c) show some representative LTu temperature profiles. Figure\ \ref{TigualT}(b) is especially noteworthy since it clearly shows that all the LTu $T(y)$-profiles sharing the same temperature values near the walls \emph{collapse} into a common curve independent of the inelasticity of the particles. Therefore, the temperature profile reached by the granular gas in the LTu flow is indistinguishable from that of an ordinary gas in the conventional Fourier flow. This surprising result is a consequence of the applicability of hydrodynamics to granular gases, even with strong inelasticity. According to the first equation in \eqref{1}, $\partial_y T^{3/2}=A\nu T^{1/2}=\text{const}$, so $T^{3/2}(y)$ is a linear function that is completely fixed by the values near the walls, regardless of the value of $\alpha$ [see inset in \ref{TigualT}(c)]. On the other hand, since $\partial T/\partial u_x=A/a(\alpha)$, the slope in the $T(u_x)$-profiles is $\alpha$-dependent, as shown in Fig.\ \ref{TigualT}(c).

\begin{figure}
  \includegraphics[width=0.75\columnwidth]{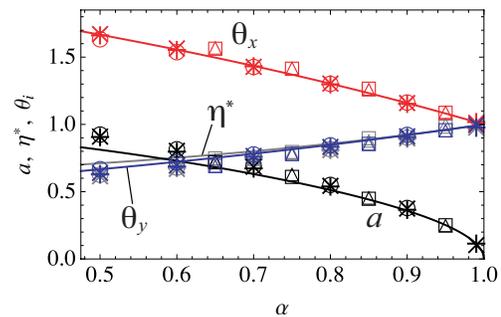}
\caption{(color online) Plot of $a(\alpha)$, $\eta^*(\alpha)$, $\theta_x(\alpha)$, and $\theta_y(\alpha)$ in the LTu flow as obtained from DSMC simulations ($h=15$) with $\Delta T=2$  ($\times$) and $\Delta  T=10$   ($+$), and from MD simulations ($h=7$) with $\Delta T=2$  ($\triangle$) and $\Delta  T=5$   ($\square$). Also DSMC data ($\bigcirc$) of the USF ($\Delta T=0$) \cite{AS05} are included. The lines represent the analytical results obtained from {the G13} approximation. Note that both in theory and simulation the values of $\theta_y$ are hardly distinguishable from those of $\eta^*$. }\label{Pfig}
\end{figure}

\begin{figure}
  \includegraphics[width=0.75\columnwidth]{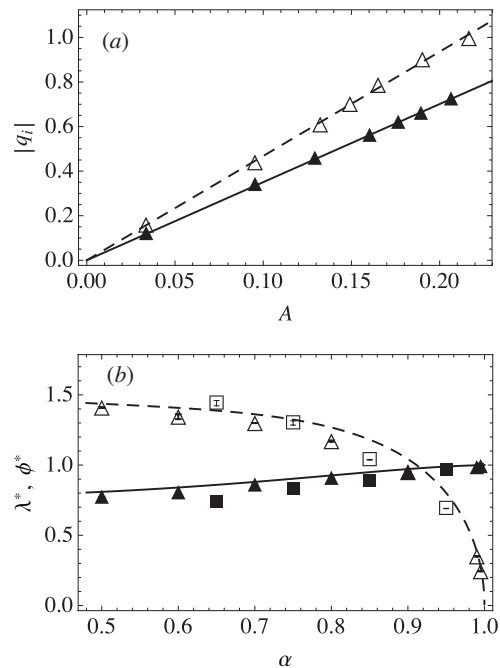}
\caption{(a) Plot of $q_x$ vs $A$ for $\alpha=0.7$  ($\triangle$) and of $-q_y$ vs $A$ for $\alpha=0.9$ ($\blacktriangle$), as {computed} from DSMC simulations.  The lines are linear fits, whose  ordinate values at $A=0$ are $1.7\times 10^{-3} $  and  $5.4\times 10^{-3}$, respectively.
(b) Heat flux coefficients $\lambda^*(\alpha)$ ($\blacktriangle$, $\blacksquare$) and $\phi^*(\alpha)$ ($\triangle$, $\square$) as  obtained from DSMC simulations (triangles) and MD simulations (squares). The lines represent the analytical results obtained from {the G13} approximation.} \label{qAfig}
\end{figure}

For each value of $\alpha$ we  have {computed} the reduced shear rate $a$ defined by the second equation of \eqref{1}, as well as the generalized transport coefficients defined   by Eqs.\ \eqref{2} and \eqref{3}. While the threshold value $\dot{\gamma}_\thr$ depends on  $\Delta T$, we have observed that, as predicted by theory, $a(\alpha)$, $\eta^*(\alpha)$, $\theta_i(\alpha)$, $\lambda^*(\alpha)$, and $\phi^*(\alpha)$ are insensitive to the choice of $\Delta T$. In particular, as shown in Fig.\ \ref{Pfig}, the reduced shear rate $a(\alpha)$ and the rheological quantities  $\eta^*(\alpha)$ and  $\theta_i(\alpha)$ are the same in the LTu-class of Couette flows (regardless of the value of $\Delta T$) as in the USF ($\Delta T=0$) \cite{AS05}, even though the boundary conditions are quite different: boundary-driven in the case of the Couette flow \cite{HH09} and Lees--Edwards periodic boundary conditions \cite{LE72} in the USF case.
Figure \ref{Pfig} also shows the close agreement between DSMC and MD results as well as the reliability of the theoretical predictions from {the G13} approximation.

Now we turn to the heat flux coefficients. One of the most striking theoretical predictions is the linear relationship between \emph{both} components of the heat flux and the thermal gradient (generalized Fourier's law), as described by Eq.\ \eqref{3}. This means that $q_i\propto A$ and this is illustrated in Fig.\ \ref{qAfig}(a) for $\alpha=0.7$ and $\alpha=0.9$. The (reduced) heat flux  transport coefficients are {evaluated} as $\lambda^*(\alpha)=-(2m/5nT)q_y/A$ and $\phi^*(\alpha)=(2m/5nT)q_x/A$, and are plotted in Fig.\ \ref{qAfig}(b). It can be observed that the streamwise component $q_x$ becomes larger in magnitude than the crosswise component $q_y$ for $\alpha\lesssim 0.9$, what represents a strong non-Newtonian effect. Interestingly, this effect, as well as the general dependence  of the transport coefficients are very well captured by our simple {G13} approximation. As happens with the rheological properties, Fig.\ \ref{qAfig}(b) shows a good agreement between DSMC and MD data for the generalized thermal conductivities.

To sum up, we have  described a special class of steady Couette flows (LTu class) in a low-density gas of inelastic hard spheres. This state encompasses the Fourier flow of elastic particles ($\partial_y T\neq 0$, $\alpha=1$) and the USF of inelastic particles ($\partial_y T= 0$, $\alpha<1$) as special cases. In this sense, LTu can be seen as a ``natural'' extension (i) of the conventional Fourier flow in ordinary gases to the realm of granular gases and  (ii) of the granular USF to states with nonzero heat flux. Therefore, the LTu uncovers a wide spectrum of inelasticities and wall temperatures within a unified framework, for both granular and elastic gases (see Fig.\ \ref{diagram}). Three complementary and independent approaches have been followed: an approximate solution based on {the G13} method, DSMC simulations of the BE, and MD simulations of a dilute system. Here, in contrast to what happens in the Couette flow for ordinary gases \cite{GS03}, {the G13} theoretical results compare surprisingly well with computer simulations, even for strong values of dissipation [see Figs.\  \ref{Pfig} and \ref{qAfig}(b)]. This paradoxical result seems to be closely tied to the condition of uniform heat flux so that, as the balance between inelastic cooling and viscous heating breaks down, {the G13} approximation is not expected to give good quantitative results, even though a hydrodynamic description still applies. The solution found here clearly shows strong non-Newtonian effects since the shear viscosity and thermal conductivity coefficients qualitatively differ from their corresponding NS values and there exist  anisotropic normal stresses and a nonzero streamwise heat flux. Despite the fact that ${\bf q}={\bf 0}$ in the USF, this state possesses intrinsic thermal conductivities in the same way as an ordinary fluid at equilibrium has intrinsic transport coefficients. Our results on the LTu class assigns a meaning to the heat flux transport coefficients of the USF, which to our knowledge have been {computed} here  for the first time.

Since MD simulations just solve Newton's equation of motion for a many-particle system, the good agreement found here between the latter method and the DSMC method shows that the existence of the class of LTu flows reported in this Letter is not an artifact of the BE, which is based on the absence of spatial and velocity correlations  (molecular chaos assumption). This fact can stimulate experiments with Couette geometry \cite{LBLG00}, where it would be possible to test whether the linear relationship between temperature and flow velocity is attainable or not.
Finally, it must be stressed that the analysis carried out here provides a nontrivial example of the existence of a hydrodynamic description for a strongly inhomogeneous state beyond the NS regime for a dilute granular gas of hard spheres.

This research has been supported by the Ministerio de Educaci\'on y Ciencia (Spain) through Grant No.\
FIS2007-60977 (partially financed by FEDER funds) and by the Junta
de Extremadura through Grant No.\ GRU09038.


\end{document}